# A Secure Communication Scheme for Corporate and Defense Community

AKM Bahalul Haque
Lecturer,
Dept. of ECE, North South University bahalul.haque@northsouth.edu
Md. Amdadul Bari
Student, Dept of ECE, North South University
Sayed Saiful Arman
Student, Dept of ECE, North South University
Farhat Tasnim Progga
Student, Dept of ECE, North South University

A Project by Cyber Threat Intelligence and Security Research Team

**Abstract**
*Security is one of the major concerns of modern communication systems. Users demand for a secure communication environment that provides privacy to the people while they are sharing messages to anyone. Privacy is a prime concern nowadays. This paper aims for providing an optimal platform for communication between the sender and the receiver. This prototype designed in the paper will provide a better anonymous path for routing messages. It will ensure ones full privacy while he or she is using this system for communication. As the proposed system provides a secure communication environment, it is supposed to be useful for the secret communication inside different governmental and non-governmental organisations. The law enforcements can use this system for any of their operations as it will encrypt and decrypt message to give a secure platform for communication.*

## 1. Introduction

Communication systems have evolved from the past in many different ways. The world has moved from analogue to digital communication and slowly revolutionised the whole concept communication system. As more and more information is being transferred everyday over the electronic media, the more it is susceptible to interception. Interceptions and tracking of communication have made enterprises and various secret government institutions vulnerable. Sensitive and confidential information might get leaked. These information can be trade secret, or matter of national security which is bad for business and country. So, a safe and secure communication system has always been a great concern.

In recent years, companies ranging from tech giants to small companies have faced data breaches. Due to data breach, user data, financial records, sensitive user information is exposed and might be used to in an illicit way to gain benefits. Identity theft has become one major concerns nowadays. So if user data is at risk, this is also a grave concern. Eavesdropping is one of the most effective communication breach. Nowadays there are a lot of equipment available for eavesdropping. In the United States a large sum of money is spent for corporate entities. Almost 300 million dollar is spent in this purpose. Every month the corporate entities are eavesdropped for various purposes. More than 8 billion dollar of loss is incurred due this eavesdropping in the US. [1]
Communication breach compromises the data confidentiality, integrity and authenticity. For example, if A and B wants to communicate with each other they will send their intended data over the communication path. If C is a potential harmful intruder and wants to know about the data, he will intercept the communication and capture the data packet. So when, B will receive the data the confidentiality will be compromised as it was already intercepted and read by C. Authenticity will also be compromised as the data was accessed by unauthorised users and also the data integrity will be compromised as B will not be sure if C has changed anything inside the data. This is also termed as man in the middle attack (MTM) [2].

## 2. Communication System

There are three components of communication which are sender, receiver and medium of communication. Sender is the source of communication. Sender sends the message through a communication path. This path may be wired or wireless. Receiver receives the message on his end.
Sender(The sending device)………...Medium……….Receiver(The recipient's device)
-------> info                                        <------ feel

A communication system comprises of the sender , receiver and the transmission medium as we can see above [3]. Internet protocols (IP) allow the communication systems to transmit data over the internet [4]. IP based communication system is one of the most popular and convenient communication systems of today's world. This communication happens using several protocols. Protocols are referred to a specific set of rules which defines how certain things within a communication system will work. Some of the protocols are given below[4].

## 3. Several Protocols for Internet Communication

Transmission Control Protocol (TCP) is a connection oriented, reliable internet protocol. Using this protocol, a connection path is build and remains connected till both the sender and receiver have finished exchanging messages [5]. The message is divided into data packets and transmitted over the network. Many of the internet applications uses TCP for the data steaming e.g. world wide web, email, remote administration etc.

User Datagram Protocol (UDP) On the other hand UDP is a connectionless protocol that is, completely opposite of the TCP [6]. UDP ensures process to process communication whereas TCP ensures host to host communication [6]. No virtual circuit path is set up before the transmission starts between sender and receiver.

Internet Control Message Protocol (ICMP) is being used for transmitting error messages to network devices, routers. ICMP is fundamentally an error-reporting protocol [4].

Hypertext Transfer Protocol (HTTP) is an application protocol that uses hyperlinks between nodes to contain text. If an user goes online and opens their web browser then indirectly they would use HTTP application protocol. User's web browser is the client of HTTP application protocol which sends request to server machines [7].

Post Office Protocol (POP) is a computer networking standard protocol which retrieves email from a remote server over TCP connection [8].

File Transfer Protocol (FTP) is a networking protocol which is used for data transmission between client and server over the network. It is used for file transfer between the client and server. In the protocol several control mechanism is used.

Internet Message Access Protocol (IMAP) is such a protocol that allows the users to access electronic mails from the mail server. This protocol performs best when user works on multiple computer networks on multiple locations.

## 4. Related Work

Scientists have tried to build an End to End Secure Messaging application on Android Device here [8]. This app uses AES-256 encryption to encrypt the message during transmission( transmission occurs using TCP) and GCM (Google Cloud Message) for routing the messages. An application for militaries and defenses which is based on XMPP protocol has been scrutinized[9]. In this application, they have proposed to use XMPP protocol for better security for the communication and encryption. On the other hand, the system architecture shows that the core servers are traceable as they are not bouncing traffics from server to server. Designing and implementation of a chat interfacing server using web-based real time have been explored in this research [10]. This system is user-friendly but the application does not use any sort of encryption because of that there will be no security for the messages. Another message sharing application has been described in this paper [11] which is based on the User Datagram Protocol(UDP). As this application is using UDP to transfer messages, so there is no surety that their message will deliver or not. Since UDP doesn't give any confirmation on successful/failed packet sending and there is no encryption and anonymity have been used for securing the information of the user from any third party. An architecture for a distributed chat application has been portrayed in this specific research [12]. In this chat system, it involves a chat server and multiple chat clients in a network that may be bandwidth constrained. Each chat client is adaptive as the system responses to instructions from the chat server so that it can operate in either a master mode or a slave mode. An adaptive, secure communication based on Unified chaotic systems with varying parameters has been explored [13]. This system includes a parameter adaptive controller which leads the response as well as get merged with the chaotic system with multiple settings so that a spontaneous communication can be built.

## 5. Solution Methodology

As it is seen in the previous part of the paper that security and privacy are one of the major concerns. An anonymous and secure route path can be established in a hostile environment [14]. When Source node needs to communicate with intended destination node, it must establish a secure and anonymous path between them in the network and then data can be safely transferred [15]. Another technique is also used which uses multi-path routing techniques to improve the effectiveness of anonymous routing [16]. Most of the time nowadays for encoding message or building this route people is using XMPP protocol. XMPP, being an Extensible Markup Language (XML) based protocol, is a widely preferred and open-standard Instant Messaging (IM) protocol. An XMPP based system is believed to be a good candidate to provide these functionalities. HTTP protocol is used to detect server-side security vulnerabilities. [17]Recent encryption algorithms play a vital role in assuring the security of Information Technology Systems and Communications, given that they can provide confidentiality, authentication, Integrity and non-repudiation. This enables the possibility to encrypt the data blocks faster. An adversary can exploit this deterministic nature of symmetric encryption algorithms by performing linear and differential cryptanalysis. Both symmetric (e.g. one time pad, AES) and asymmetric (e.g. RSA and ElGamal cryptosystems) key encryption algorithms can be exploited to accomplish decryption mix nets. The new design of a high availability system, with the use of High Availability Proxy and the load balancing with Domain Name Service, both using the round robin algorithm, develop a new system easy to manage, flexible and scalable. For message routing this system is using XMPP. XMPP used for instant messaging, presence, multi-party chat, voice and video calls, collaboration, lightweight middleware, content syndication, and generalized routing of XML data. XMPP is the exchange of small, structured chunks of information (in XML). XMPP is like HTTP a client server protocol, but the difference between these two is XMPP allowing either side to send data to the other asynchronously. XMPP Connections are very powerful and data is not pulled it is pushed. In XMPP server if we want, we can add functionality in server by writing hooks in server. In this system Advanced Encryption Standard (AES) 256 bits has been used as the specification of encryption. There are no successful security attacks against AES have been casted yet [1]. If the security strength of AES is considered against brute force attack, mathematical attack, timing attack the rate of successful attacks are below average [1].

A secure communication system should have some very specific requirements, like no third party involvements, ensuring data protection, promoting anonymity, using hard routing methods and high level encryption and decryption methods in the system. Keeping all these attribute under consideration we have tried to develop a secure communication system including some of the attributes mentioned here.

## 6. Encryption

Encryption method process means that it can encode messages in a way that only the authorized people can access it. If one user sends message to another user by using encryption method then nobody else can encode the message expect the desired person. For this there are two popular encryption systems Symmetric and Asymmetric encryptions are here to help.

**6.1 Symmetric encryption** – In the symmetric (privatekey) one, encryption and decryption are performed under a key shared by the sender and receiver [18] . Symmetric encryption allows a party to outsource the storage of his data to another party in a private manner, while maintaining the ability to selectively search over it [19].

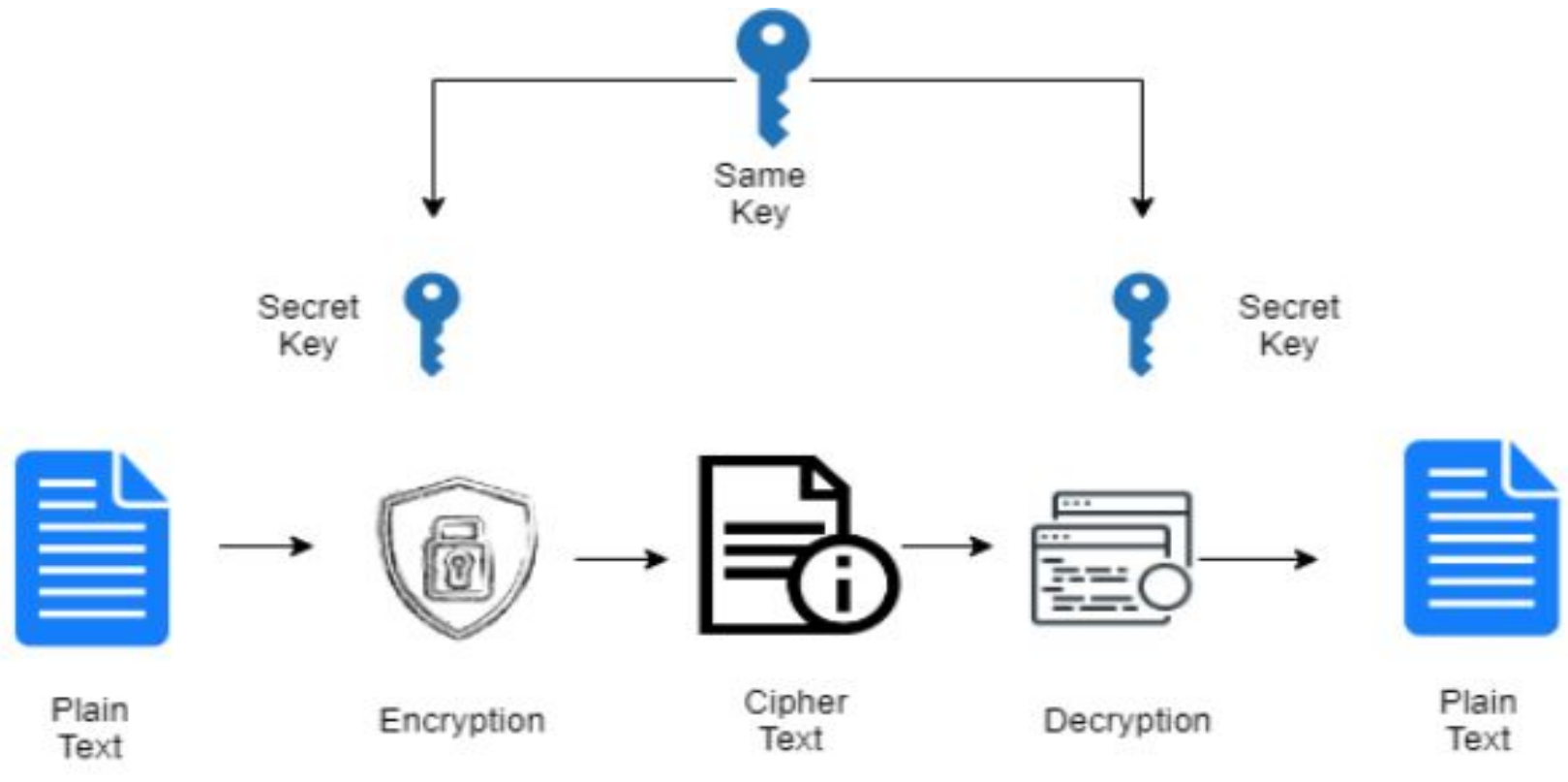

Symmetric Encryption

**6.2 Asymmetric Encryption** – An asymmetric encryption scheme is usually employed only for distributing a secret-key of a symmetric encryption scheme for message encryption. Public key is used for encrypting our data or message and private key is used for decrypting the encrypted message. In the asymmetric (public-key) setting the sender has some public data and the receiver absorbs some corresponding secret personal data [20]. Look at the following diagram for understanding how Asymmetric Encryption works.

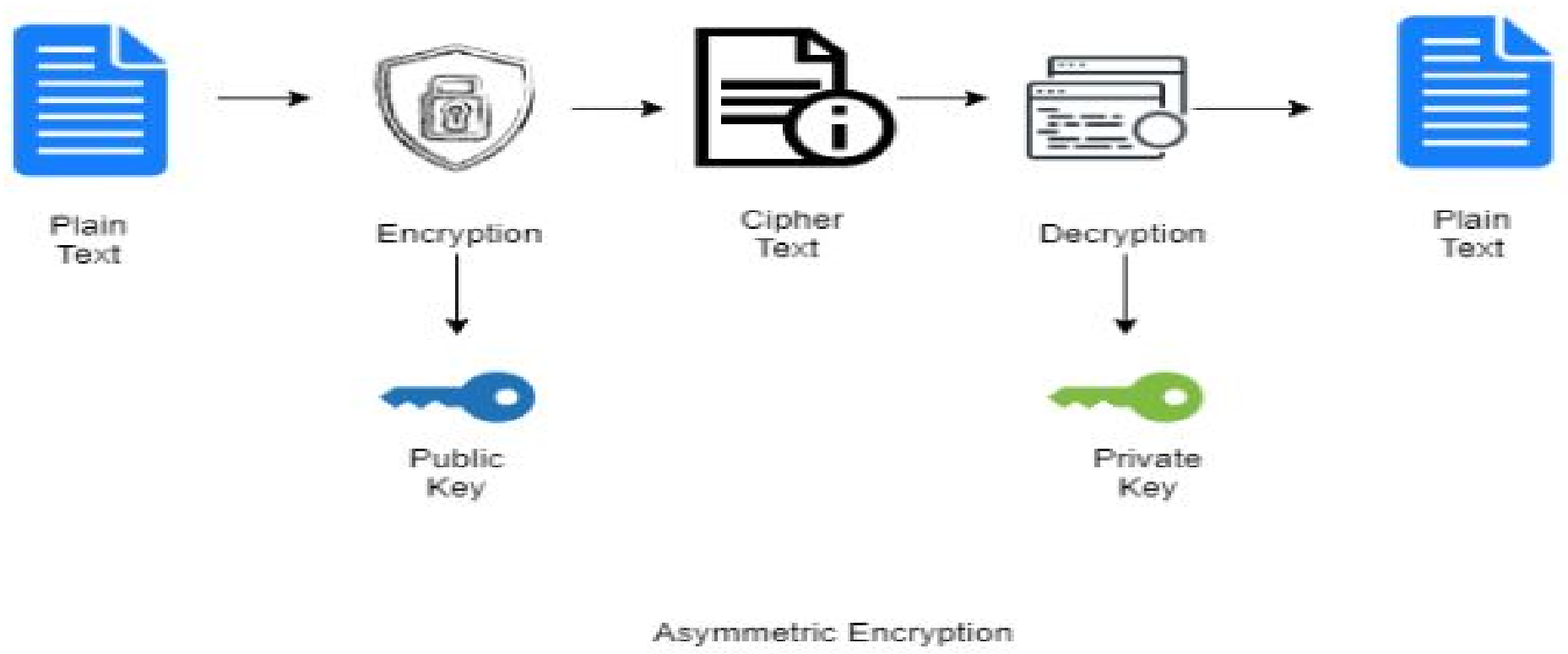

## 7. Design

The system consists of XMPP, HTTP protocols, HAproxy and the other equipments of the system have been aligned in Raspberry Pi to make it a device not just an application. As it will focus on the security of law enforcements data so it will be better for this system to be a device not an application. The system architecture of this device has been described later on this report. It will emphasize how the system will work during its interfaces. The system is designed for the security of confidential information that are shared among the law enforcements. By being anonymous it will give them the privacy of not getting vulnerable.

## 8. System Architecture

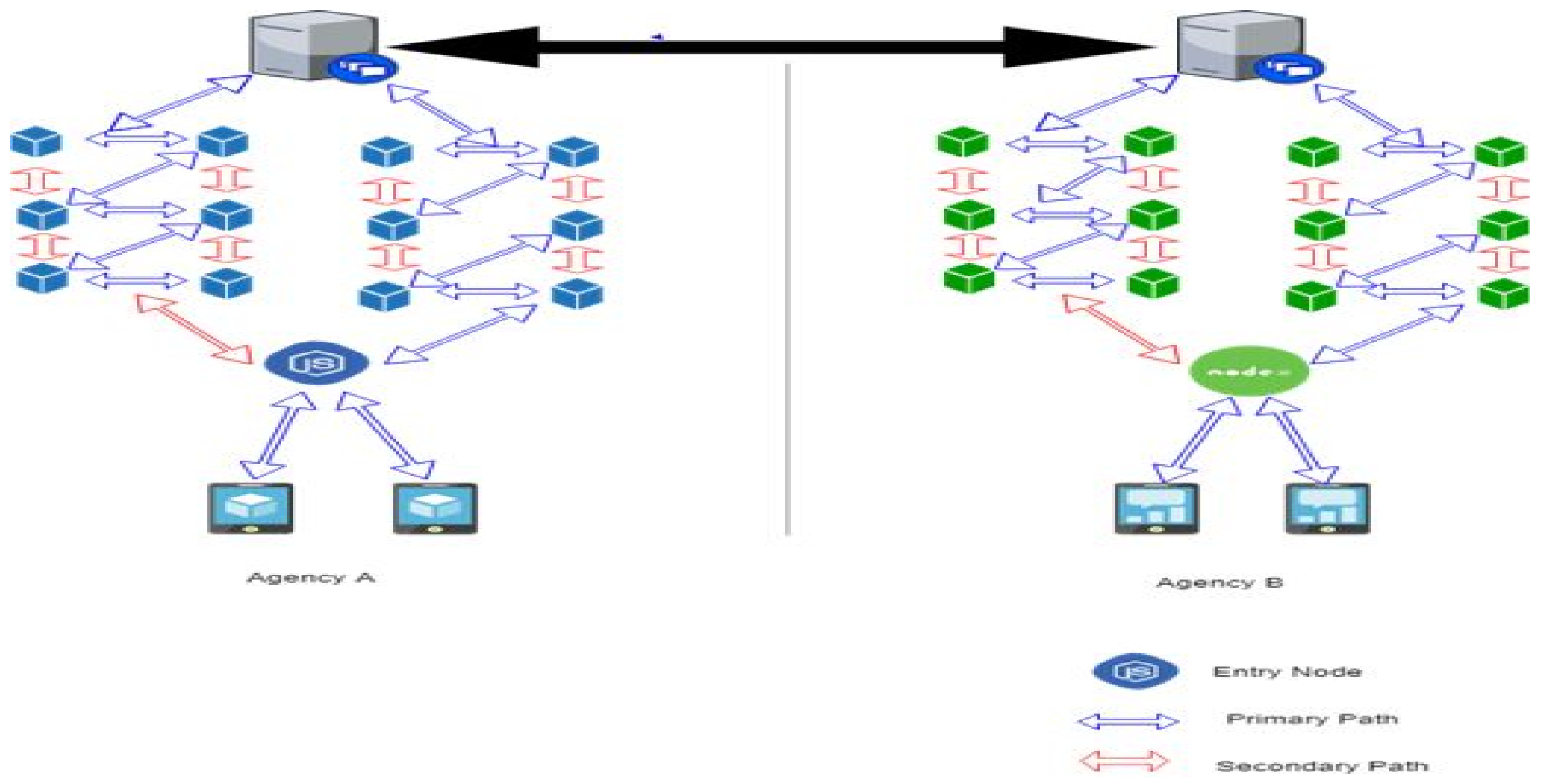

Figure 1 : Figure 3 System Architecture

In this system architecture diagram there may happen four types of communication -

-Agency A to Agency A communication.

-Agency A to Agency B communication.

-Agency B to Agency B communication.

-Agency B to Agency A communication.

If Agency A wants to communicate with someone from agency A the source will knock the destination by using device. When the source device sends any message to destination device for communicating firstly the message will go to the server. Here, it used Openfire server and the message will get encrypted by using the public key of the server. After that, it will go to the entry node. This entry node is working as a load balancer and firewall. for load balancing it will be using HAProxy and it follows a random path of nodes by using Round Robin Algorithm and it ends in agency A's server which will make anonymous.

After that, the private key of the server will decrypt the message so that if it's required to get Monitored and this will make the procedure more strong. The message will then get encrypted by The destination device's public key and it will arrive on the destination by going through another random path of nodes. The destination device will decrypt the message by its private key as soon as it gets the message from the node as well as the destination user will see the popped up message on his or her device. The same thing will happen when a person of Agency B wants to Communicate with anyone from Agency B.

Agency A to B communication process is almost same first agency A will complete its process to agency A server then agency A server will decrypt the message and see the destination and encrypted with server B public key and pass the message to agency B server. Then server B will see the destination by decrypted the message with server B private key and again encrypted the message using destinations device public key of the server. Then agency B will do its process and go to the destination device then destination device and decrypt the message by its private key. As there is two agencies communicating with each other so there are will be two servers working at the same time so both servers will have the record of this conversation.

## 9. User Interface

In this Section, how our application works. Some attachments are given below of the interfaces of the system. It emphasizes how the system works during particular interfaces.

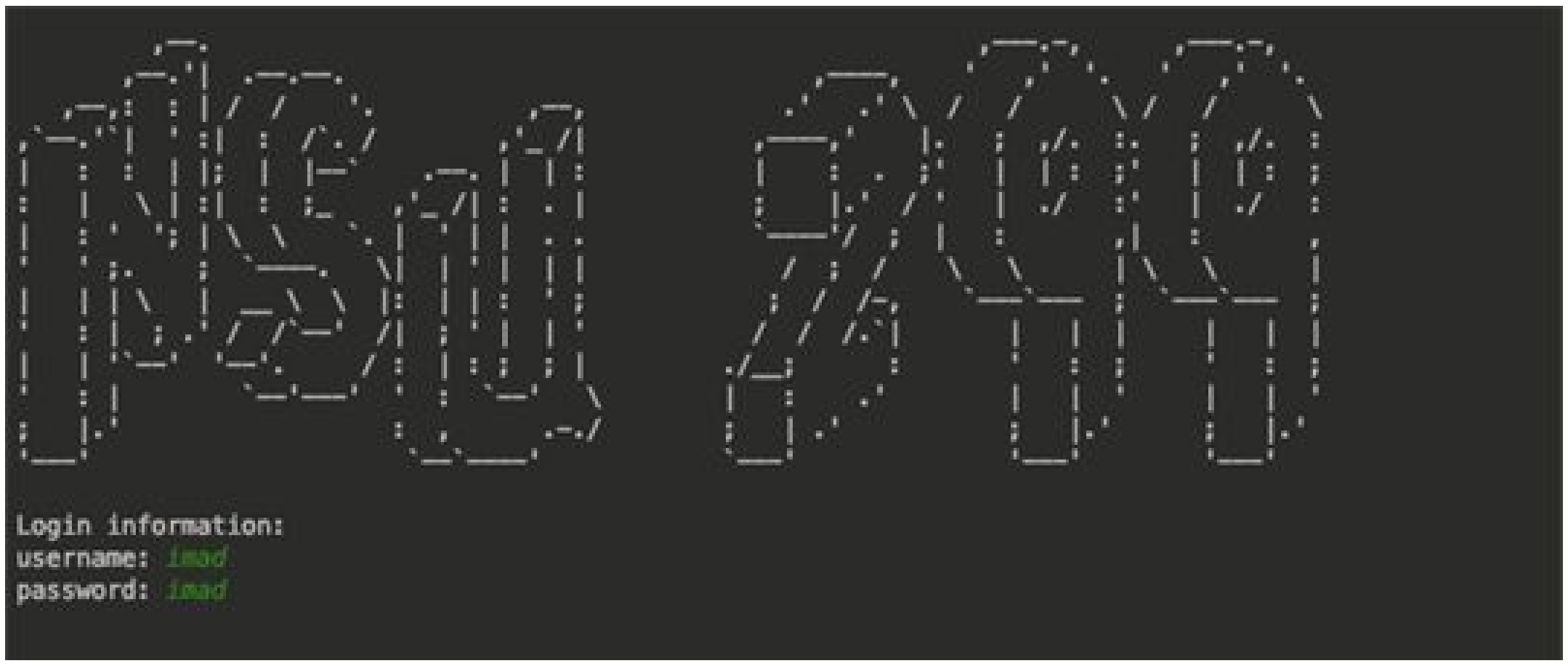

**Figure 4 Logo and Login Information**

Above figure is the command line interface. When one open the system the logo of that system will pop up. Then, the authentication part comes and for that one has to log in to the system which contains username and password for verification or register.

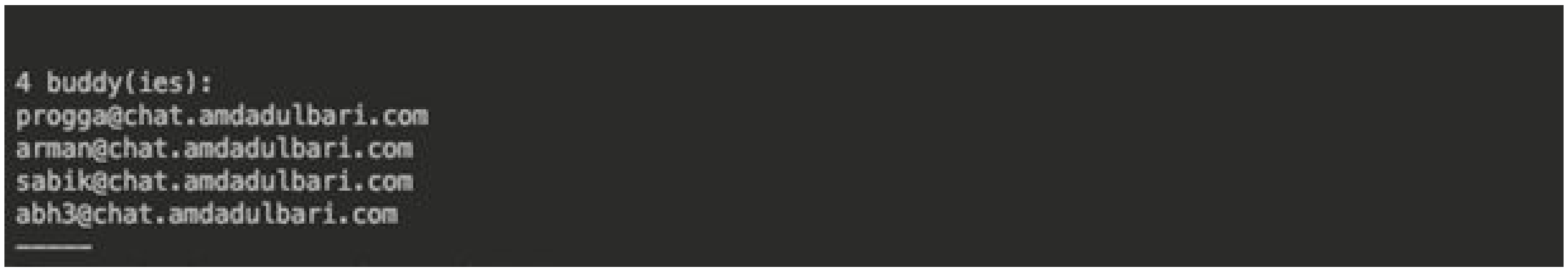

**Figure 5 Friend list/Buddy list**

After logging in user will see their friend list. In this section they will find their friend list as a buddy list. In list one can only see the people whom they have added via email. They also can choose the person with whom they want to connect instantly.

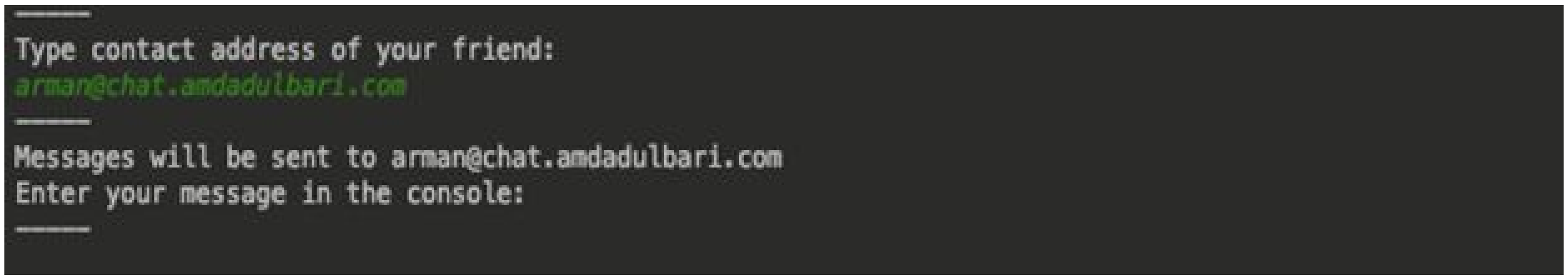

Figure 6 Chatting procedure

Here, one has to enter the contact address of their friends with whom they wanted to connect. Entering the contact address means the message will be sent only this contact nobody else will be able to see it. After inputting their desired contact message this will be sent to their friends and they will see the message going too their recipient friends.

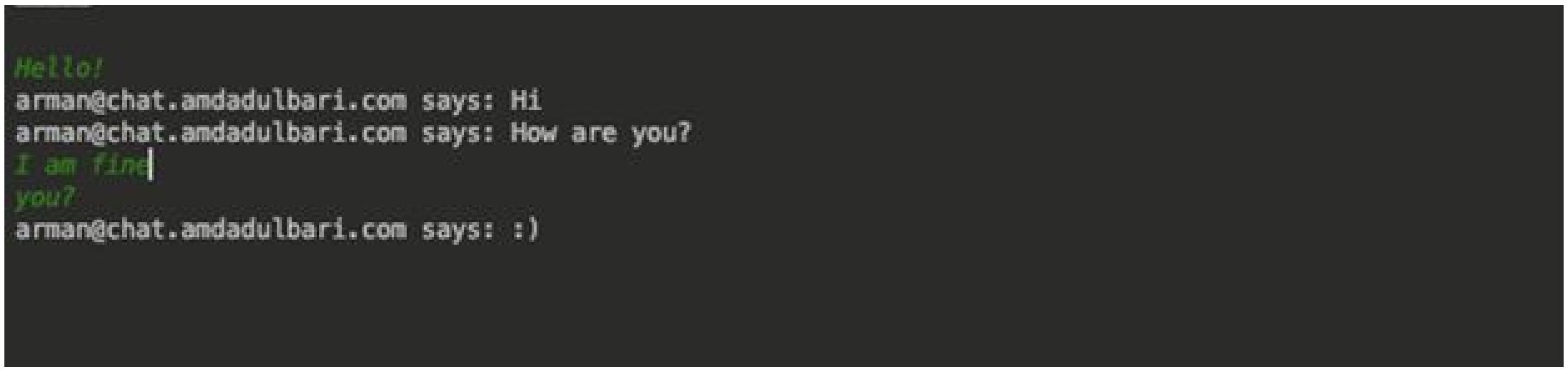

**Figure 7 Conversation**

If user is sends “Hello” to his friend. When user’s friend sees the messages they will get back to them by replying to their messages which has been shown in the upper interface.

## 10. Result

After assembling all the tools the system will work as a secure communication device. If the interfaces are taken as result of the system it can be considered that the system is encrypting the messages properly and so do the routing path to be the anonymous one.

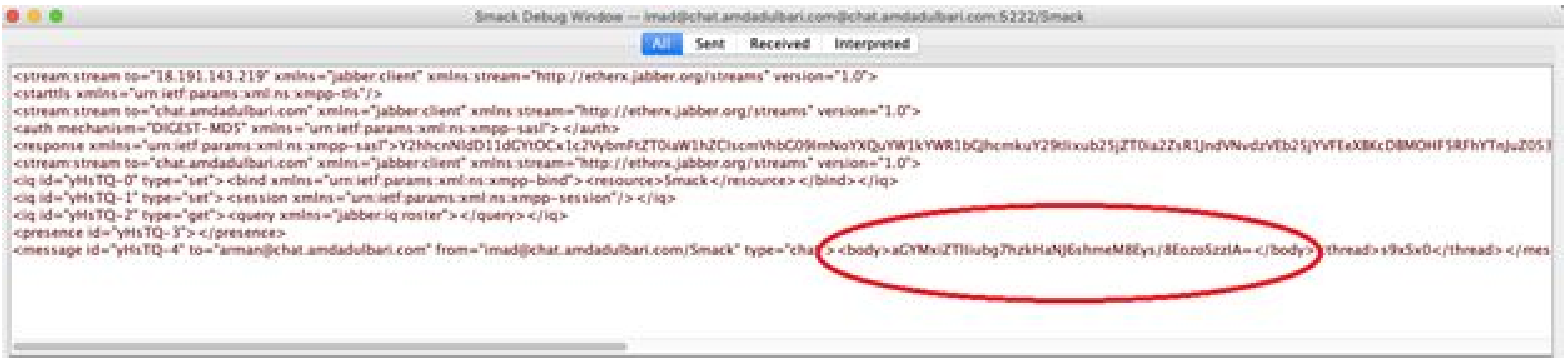

**Figure 8 Debug Window**

The above figure is the debug window of SMACK API. This API is used for implementing XMPP protocol.In the figure 8 the red marked portion contains the message body. The message body is fully encrypted. For the encryption its using both asymmetric (RSA 2048) and symmetric (AES 256) encryption.  This two encryption are very strong and latest encryption algorithm till date for secure data transmission.

## 11. Scopes

In future some features can be added to the system. As the system contains Command Line Interface so there is nothing much to be added. It is totally different from graphical user interface. If it used GUI there would be several options to add many features. However, for the CLI feature voice message, audio call, video call, push notification and many other features cannot be added to the system. So, there is not much features to add in future. Besides that, some features like file sharing, photos/videos sharing can be added. So that, the user of this application especially business organizations and corporate agencies can rely on this application and send each other important and confidential files, photos and videos by using our secure application. As this application is secured so it will also protect those files, photos and videos in the same way. Though sending videos and photos might take long time to encrypt and also it will be needed high bandwidth but we are adding these features for business organizations and corporate agencies and they obviously use high bandwidth connection. So, it might take little time but not take much longer time to encrypt photos or even videos. Other than that, some development can be done on updating the applications authentication. For that, adding fingerprint login and retina login can be very useful at the same time it will ensure more security for the users. Adding this authentication will make sure nobody else is using the user's ID. Because without user's valid fingerprint or retina nobody will access his/her account. So, adding this authentication will make this application's security much stronger than before.

Nowadays privacy and anonymity is a big concern for us. Online activities are being tracked on the internet [21]. Each and every footprint that we have in the world can be used and tracked back to us. For this reason information sharing through a secure communication channel is a concern for common people also.

## 12. Conclusion

As security is a vulnerable issue the communication system which is described earlier can be certainly considered as a secure communication media for law enforcements to protect their confidential data. Law enforcements can use this system for their own use to protect their data from outsiders. On the other hand business corporations and other agencies can also use this system so that in a competitive market they can protect their data . Security can be maintained through this communication system. Secure communication system implies a system which effectuates all the security requirements of the system and the above system considered to fulfill the requirements. It can be used for the welfare of the mankind as it will protect the privacy of its users.